\documentclass{article}
\usepackage[super,compress]{cite}
\usepackage{graphicx}
\usepackage{hyperref}

\begin{document}

\markboth{C. Cuesta \em{et al.}}
{Analysis of the $^{40}$K contamination in NaI(Tl) crystals from different providers in the ANAIS project}

\title{ANALYSIS OF THE $^{40}$K CONTAMINATION IN NAI(TL) CRYSTALS FROM DIFFERENT PROVIDERS IN THE FRAME OF THE ANAIS PROJECT.}

\maketitle
\begin{center}
\author{C. CUESTA\footnote{Presently at Center for Experimental Nuclear Physics and Astrophysics, and
Department of Physics, University of Washington, Seattle, WA, US, ccuesta@uw.edu}, J.~AMAR\'E, S.~CEBRI\'AN, E.~GARC\'IA, C.~GINESTRA, M.~MART\'INEZ\footnote{Fundaci\'{o}n ARAID, Mar\'{\i}a de Luna 11, Edificio CEEI Arag\'{o}n, 50018 Zaragoza, Spain}, \\
M.A.~OLIV\'AN, Y.~ORTIGOZA, A.~ORTIZ~DE~SOL\'ORZANO, C.~POBES\footnote{Presently at Instituto de Ciencia de Materiales de Arag\'on, University of Zaragoza - CSIC}, J.~PUIMED\'ON,  \\
M.L.~SARSA, J.A.~VILLAR, P. VILLAR.}
\end{center}

\begin{center}
Laboratorio de F\'{\i}sica Nuclear y Astropart\'{\i}culas, Universidad de Zaragoza, C/~Pedro Cerbuna 12, 50009 Zaragoza, SPAIN\\
Laboratorio Subterr\'aneo de Canfranc, Paseo de los Ayerbe s.n., 22880 Canfranc Estaci\'on, Huesca, SPAIN
\end{center}


\begin{abstract}
NaI(Tl) large crystals are applied in the search for galactic dark matter particles through their elastic scattering off the target nuclei in the detector by measuring the scintillation signal produced. However, energies deposited in the form of nuclear recoils are small, which added to the low efficiency to convert that energy into scintillation, makes that events at or very near the energy threshold, attributed either to radioactive backgrounds or to spurious noise (non-bulk NaI(Tl) scintillation events), can compromise the sensitivity goals of such an experiment. DAMA/LIBRA experiment, using 250\,kg NaI(Tl) target, reported first evidence of the presence of an annual modulation in the detection rate compatible with that expected for a dark matter signal just in the region below 6\,keVee (electron equivalent energy). In the frame of the ANAIS (Annual modulation with NaI Scintillators) dark matter search project a large and long effort has been carried out in order to understand the origin of events at very low energy in large sodium iodide detectors and develop convenient filters to reject those non attributable to scintillation in the bulk NaI(Tl) crystal. $^{40}K$ is probably the most relevant radioactive contaminant in the bulk for NaI(Tl) detectors because of its important contribution to the background at very low energy. ANAIS goal is to achieve levels at or below 20\,ppb natural potassium. In this paper we will report on our effort to determine the $^{40}$K contamination in several NaI(Tl) crystals, by measuring in coincidence between two (or more) of them. Results obtained for the $^{40}K$ content of crystals from different providers will be compared and prospects of the ANAIS dark matter search experiment will be briefly reviewed.
\end{abstract}


\section{Introduction}	
\label{sec:Intro}

ANAIS project aims at the study of the annual modulation signal attributed to galactic dark matter particles using 250\,kg NaI(Tl) scintillators at the Canfranc Underground Laboratory (LSC), in Spain. NaI(Tl) large crystals have been applied for a long time in the search for galactic dark matter particles through their elastic scattering off the target nuclei in the detector by measuring the weak scintillation signal produced \cite{DAMA,LIBRA,DM32,JAPAN1,PSD_Gerbier,UKDMS,JAPAN2,DM-ice}. However, energies deposited in the form of nuclear recoils are small, which added to the low efficiency to convert that energy into scintillation, makes that events at or very near the energy threshold, attributed either to radioactive backgrounds or to spurious noise (non-bulk NaI(Tl) scintillation events), can compromise the sensitivity goals of such an experiment. DAMA experiment, at the Laboratori Nazionali del Gran Sasso, in Italy, and using 100\,kg NaI(Tl) target, reported first evidence of the presence of an annual modulation in the detection rate compatible with that expected for a dark matter signal just in the region below 6~keVee (electron equivalent energy) with a high statistical significance \cite{DAMA}. This signal was further confirmed by LIBRA experiment, using 250\,kg of more radiopure NaI(Tl) detectors \cite{LIBRA}. Using the same target than DAMA/LIBRA experiment, which accumulates by now fourteen annual cycles, makes possible for ANAIS to confirm DAMA/LIBRA results in a model-independent way. To achieve such a goal ANAIS detectors should be as good (or better) as (than) those of DAMA/LIBRA in terms of energy threshold and radioactive background below 10 keVee (electron equivalent energy).
In this paper we will present some of the past and recent efforts to determine and reduce the background related to $^{40}K$ contamination in several prototypes, as well as to determine the achievable threshold profiting from the low energy events population, conveniently tagged, that such a contamination provides.
We will start by presenting the ANAIS project and the different crystals and experimental set-ups studied; then, we will move to understand the importance of the background due to $^{40}K$ contamination and the technique used to determine its content in the different crystals, in particular, in ANAIS-25 modules, as well as the results derived.

\section{ANAIS project and NaI(Tl) crystals studied}
\label{sec:ANAIS}

\begin {figure}[h!]
\centerline{\includegraphics[width=0.35\textwidth]{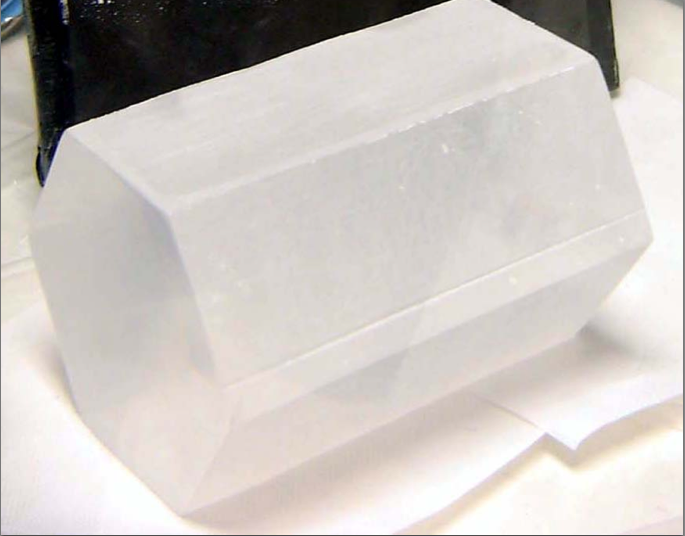}
\includegraphics[width=0.2\textwidth]{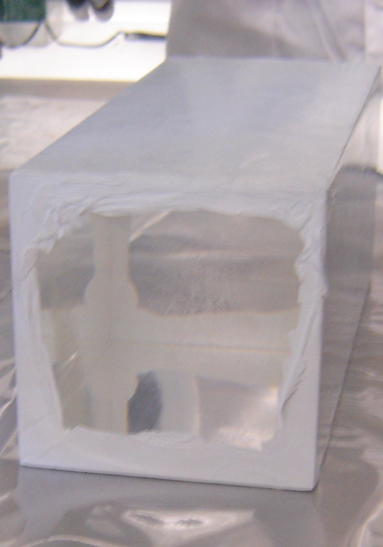}
\includegraphics[width=0.35\textwidth]{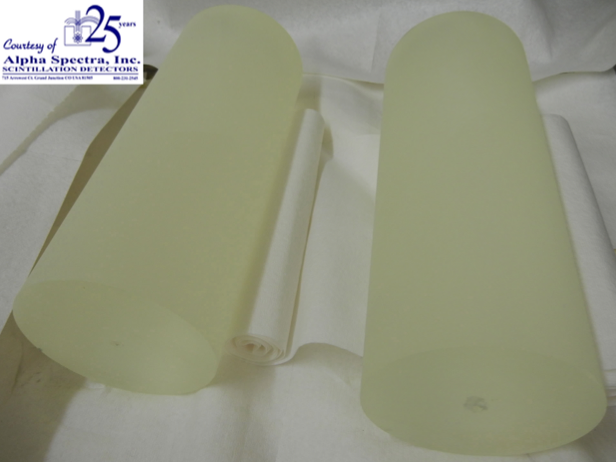}}
\caption{Different NaI(Tl) crystals studied at the LSC in order to determine their potassium content by the coincidence technique: 10.7\,kg crystal from BICRON (PIII) - left, 9.7\,kg crystal from Saint Gobain (ANAIS-0) - center, and 12.5\,kg crystals from Alpha Spectra (ANAIS-25) - right.}
\label{fig:crystals}
\end{figure}

ANAIS project was conceived to use 10 hexagonal NaI(Tl) crystals of 10.7\,kg each made by BICRON (now Saint Gobain) in the eighties. These detectors were part of those used in an experiment which looked for the $^{76}Ge$ $\beta\beta$ decay to the first excited state at the Modane Underground Laboratory first~\cite{modanedobleb2}, and then at the LSC~\cite{moralesdobleb2}. They have been stored underground since the late eighties. After that, the NaI32 experiment searched for dark matter at the LSC with 3 of these detectors (amounting 32.1\,kg) and accumulating two years of data taking. Bounds on WIMP masses and cross-sections were derived from the absence of positive hints, both in the usual analysis of the total rate, and in a pioneer modulation analysis~\cite{DM32}. Then, one of the NaI crystals was chosen to be further studied and modified: it was used to build ANAIS Prototype~I~\cite{tesisSusana}, and after decoupling the PMT and removing the original stainless-steel encapsulation it was used in ANAIS Prototypes~II~\cite{tesisMaria} and~III (see Fig.\,\ref{fig:crystals}, left). In Prototype~II a copper box was used to allow testing easily different light guides geometries and lengths, optical couplings and reflector/diffuser materials, whereas in Prototype~III (PIII) a tight copper encapsulation was designed and PMTs and optionally light guides were coupled in a second step. The $^{40}K$ content of these crystals (see section \ref{sec:resultsK}) was  too high to allow their use in a dark matter search experiment, which implied a significant change in the ANAIS experiment time-line.

The ANAIS collaboration started to look for new NaI(Tl) crystals with very low content in potassium (less than 20\,ppb). Saint Gobain \cite{SaintGobain} was contacted as the first option to produce the new crystals, as it had low radioactive background state-of-the-art NaI(Tl) detector technology: they had previously built DAMA/LIBRA detectors with similar background requirements. A 9.7\,kg NaI(Tl) crystal had been bought previously to that company to be tested as first step in this direction. This crystal  (see Fig.\,\ref{fig:crystals}, center) was used to build the ANAIS-0 module. The crystal was encapsulated at the University of Zaragoza in ETP (Electrolytic Tough Pitch) copper, closing tightly the detector, and using two synthetic quartz windows to get the light out to the photomultiplier tubes. A Teflon sheet as diffuser and a Vikuiti\texttrademark reflector layer wrapped the crystal to increase light collection efficiency. ANAIS-0 module was designed to characterize and understand ANAIS background at low energy, optimize NaI scintillation events selection, fix the calibration method and test the electronics. ANAIS-0 module took data in different configurations: with or without 10\,cm light guides and using different PMT models \cite{TesisClara}. A detailed background study of ANAIS-0 module in the different configurations tested has been recently published~\cite{ANAISbkg}. Levels of $^{40}K$ of that crystal were not enough for ANAIS requirements, as will be shown in section \ref{sec:resultsK_a}. Unfortunately, the contacts with Saint Gobain to develop more radiopure crystals were not successful.

The ANAIS experiment had then to be redefined and, in order to reach the experimental goals, the final proposal consisted of 250\,kg of ultrapure NaI(Tl) crystals to study the expected annual modulation in the galactic dark matter signal. An ultrapure powder provider and manufacturer of the ultra-low-background crystals had to be searched for. A NaI powder having a potassium level under the limit of the analytical techniques used for the radioactivity screening (less than 90\,ppb) was found, and then, two crystals were grown by Alpha Spectra \cite{Alpha Spectra} to determine, afterwards, more precisely the potassium content using the coincidence technique presented in section\,\ref{sec:40K}. These two NaI(Tl) crystals (12.5\,kg each) form the ANAIS-25 setup, and are named detector 0 (D0) and 1 (D1), respectively (see Fig.\,\ref{fig:crystals}, right). The crystals were encapsulated in OFHC copper with two synthetic quartz windows allowing the PMTs coupling in a second step, to be done at LSC clean room. Only white Teflon was used as light diffuser, wrapping the crystal, inside the copper encapsulation. After the determination of the corresponding potassium level (presented in section\,\ref{sec:resultsK_b}), these detectors are still taking data at the LSC, remaining as main goal an overall background assessment\,\cite{TesisClara}.

\section{Determination of the $^{40}$K bulk content in NaI(Tl) by measuring in coincidence}
\label{sec:40K}

$^{40}K$ is probably the most relevant radioactive contaminant in the bulk for NaI(Tl) detectors. It is specially important its contribution to the background at very low energy, as it will be shown below. Because of that, a very good knowledge of such a contamination is required in order to properly estimate the sensitivity prospects achievable in the frame of the ANAIS project.

\begin{figure}[h!]
\centerline{\includegraphics[width=0.60\textwidth]{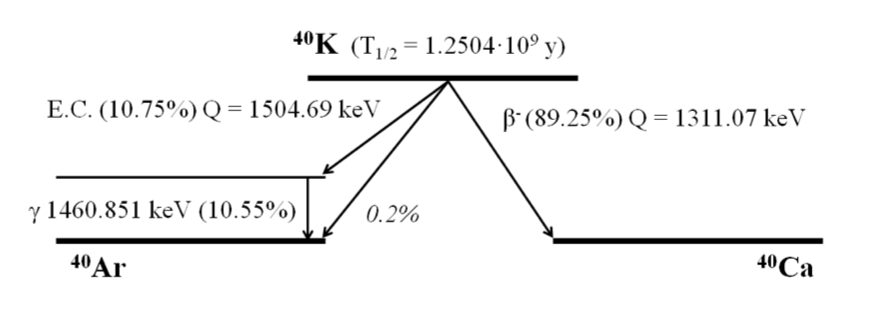}}
\caption{Simplified $^{40}$K decay diagram. Data have been taken from \cite{nucleide}. More details about branching ratios and decay channels are given in Table \ref{tab:decayK}.} \label{fig:decayK}
\end{figure}

The technique used to estimate the $^{40}K$ activity in the bulk of the NaI(Tl) crystal is the measurement in coincidence: one detector measures the energy released by the X-ray/Auger electrons emissions of argon, amounting a total energy release of 3.2\,keV, following the K-shell EC of $^{40}K$, while another detector measures the high energy gamma at 1460.8\,keV, escaping from the former, and being fully absorbed in the latter. Besides a few accidental coincidences, the 3.2\,keV peak can be clearly observed in all the studied crystals. From the measured coincidence rates and the corresponding efficiencies estimated with Geant4 package, the $^{40}K$ activity in the crystal can be deduced. In Figure \ref{fig:decayK} a simplified $^{40}K$ decay diagram is shown and more details on the $^{40}K$ decay channels can be found in Table \ref{tab:decayK}.

\begin{table}[h!]
\begin{center}
\caption{Information about the different decay channels available for the $^{40}$K decay: EC to the first excited state ($\epsilon_{0,1}$), EC to the ground state ($\epsilon_{0,0}$), $\beta^+$, and  $\beta^-$. Only EC to the first excited state is followed by a gamma emission. In the case of the EC channels the relative probabilities of K/L/M capture are also given: P$_K$, P$_L$, and P$_M$.}
\vspace{0.3cm}
{\begin{tabular}{@{}cccccc@{}} 
\hline
 & Energy (keV) & Probability (\%) & $P_K$ & $P_L$ & $P_M$ \\
 \hline
$\epsilon_{0,1}$ & \hphantom{00}44.0\hphantom{0} $\pm$ 0.3\hphantom{0} & 10.55 $\pm$ 0.11 & 0.763 & 0.209 & 0.027 \\
$\epsilon_{0,0}$ & 1504.69 $\pm$ 0.19 & \hphantom{0}0.2\hphantom{0} $\pm$ 0.1\hphantom{0} & 0.880 & 0.086 & 0.013 \\
$\beta^+_{0,0}$ & \hphantom{0}489.3\hphantom{0} $\pm$ 0.3\hphantom{0} & \hphantom{0}(1.00 $\pm$ 0.12) $10^{-3}$ &  &  &  \\
$\beta^-_{0,0}$ & 1311.07 $\pm$ 0.11 & 89.25 $\pm$ 0.17 &  & & \\
\hline
\end{tabular}
\label{tab:decayK}}
\end{center}
\end{table}

Concerning the contribution to the background at low energy from the decay of this isotope, it is precisely the X-ray/Auger electrons emission of argon after K-shell EC in $^{40}K$, amounting a total energy release of 3.2\,keV, which is more dangerous. A part of these events is moved to higher energies by the simultaneous interaction of the high energy gamma in the same detector, another part is vetoed by an interaction of the high energy gamma in another detector, however, a residual peak at 3.2\,keV jeopardizes the search for dark matter if $^{40}K$ content is not at a few ppb level. At this point, it is worth noticing that the rate of the direct decay of $^{40}K$ to the ground state of $^{40}Ar$ through EC (at the 0.2\% level) has not been yet experimentally measured, and hence, the uncertainty on the corresponding branching ratio is really high \cite{Pradler}. This decay is important for the final ANAIS background because those events can not be rejected by the coincidence method. At this moment, and up to our knowledge, there is not a feasible way to determine more precisely such a branching ratio.

\section{Brief description of the experimental set-ups}
\label{sec:expsetup}

\begin {figure}[h!]
\centerline{\includegraphics[height=0.195\textheight]{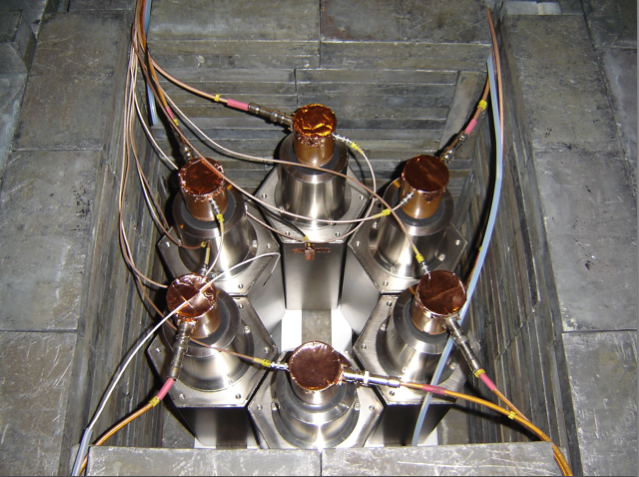}
\includegraphics[height=0.195\textheight]{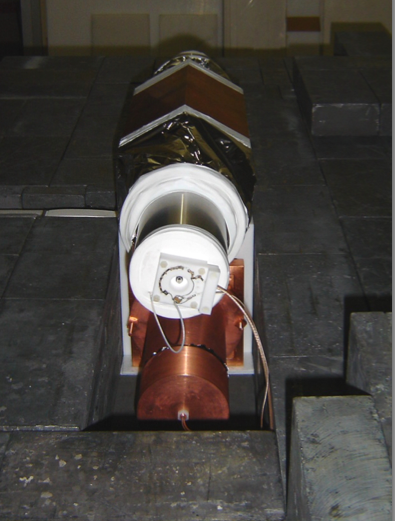}
\includegraphics[height=0.195\textheight]{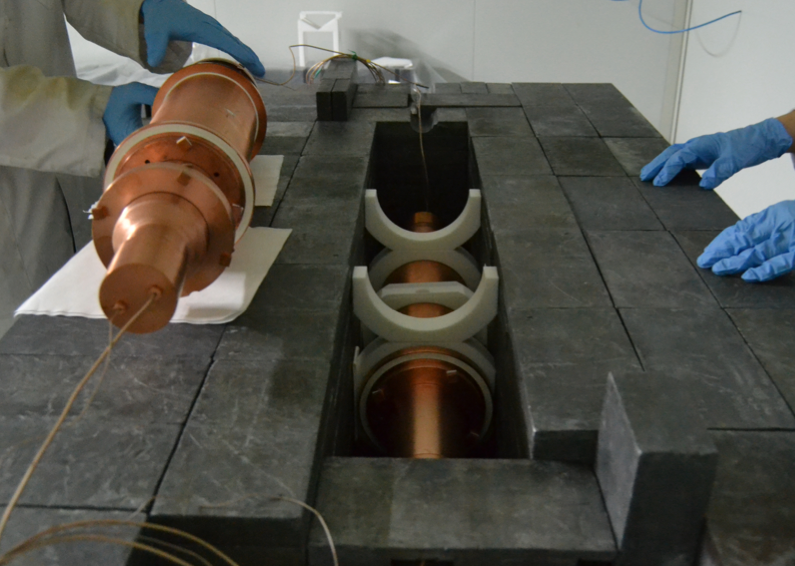}}
\caption{Different set-ups operated at LSC to measure the $^{40}K$ content of the available NaI(Tl) crystals and described in more detail in the text. Left: BICRON set-up, center: ANAIS-0 and PIII set-up, and right: ANAIS-25 set-up. Complementary information is also given in Table~\ref{tab:setups}}
\label{fig:setups}
\end {figure}

Different dedicated setups were operated at the LSC with the objective of determining the $^{40}K$ content of all the NaI(Tl) available crystals. Pictures of those crystals have been shown in Fig.~\ref{fig:crystals} and their main features are summarized in Table~\ref{tab:crystals}. We describe below briefly the most relevant aspects of the different experimental set-ups, corresponding pictures are shown in Fig.~\ref{fig:setups}, in which the coincidence measurements were carried out. All the set-ups have been operated at the LSC facilities (under 2450\,m.w.e.), inside a shielding consisting of 10\,cm archaeological lead plus 20\,cm low activity lead, all enclosed in a PVC box tightly closed and continuously flushed with boil-off nitrogen.

\begin{table}[h!]
\begin{center}
\caption{Main features of the different crystals studied in this work: set-up where they have been operated, manufacturer, mass, shape and dimensions$^*$.}
\vspace{0.3cm}
{\begin{tabular}{@{}cccccc@{}} 
\hline
Set-up& Detector		& Manufacturer  & Mass      &   Shape               & Dimensions\\ 	
\hline	
1  & EP054, EP055, EP056,   & BICRON        & 10.7\,kg  & hexagonal prism       & 15.94\,cm\,$\times$\,20.32\,cm\\
   & EP057, EP058, EP059, & & \\
   & EL214, EM301, EL604, & & \\
   & EL603, EL607 & & \\
2 &  PIII (EL607) & BICRON        & 10.7\,kg  & hexagonal prism       & 15.94\,cm\,$\times$\,20.32\,cm\\
2 &  ANAIS-0      & Saint Gobain  & 9.6\,kg   & parallelepiped prism  & 10.16\,$\times$\,10.16\,$\times$\,25.40\,cm$^{3}$\\
3   &  ANAIS-25 D0  & Alpha Spectra & 12.5\,kg  & cylinder     & 4.75" ($\Phi$)\,$\times$\,11.75"\\
 &    ANAIS-25 D1  &   &   &   &  \\
\hline
\multicolumn{6}{p{\textwidth}}{* For the cylinders are given diameter and length, and for the hexagonal prism, distance between opposite vertices in the hexagonal face and length.}
\end{tabular}\label{tab:crystals}}
\end{center}
\end{table}

 \begin{enumerate}

    \item \textbf{BICRON set-up.} In this set-up the BICRON detectors were measured without modification with respect to their original assembly. Because of that, only one PMT per crystal was recording the scintillation signal. In order to mostly profit from the large number of crystals available, six and seven crystal units were measured simultaneously in two runs (labeled as {\it a} and {\it b} in Table~\ref{tab:setups}). Although using only one PMT signal was quite limiting to lower the experimental threshold, the tagging with the high energy gamma from potassium allowed to distinguish clearly the 3.2\,keV peak in all the crystals and to derive the results shown in section\,\ref{sec:resultsK}. A picture of the setup is shown in Fig.~\ref{fig:setups}, left.

    \item \textbf{ANAIS-0 and Prototype III set-up.} The estimate of the $^{40}K$ bulk content of the ANAIS-0 crystal, was done by measuring in coincidence with the previously studied Prototype III in a dedicated setup, shown in Fig.~\ref{fig:setups}, center. Measurements were carried out in several phases, see Table~\ref{tab:setups}, and although Phase III had another purpose~\cite{ANAISom}, the corresponding data will be revised too in section \ref{sec:resultsK_a}. Data have been analyzed separately for each phase and $^{40}K$ bulk content of the PIII has been also determined as cross-check.

    \item \textbf{ANAIS-25 set-up}. The main goal of the ANAIS-25 set-up was precisely the determination of the potassium content of the new crystals. It consists of two cylindrical 12.5\,kg NaI(Tl) detectors grown with ultrapure NaI powder ($<$90\,ppb potassium at 95\% CL according to results from HPGe spectrometry screening carried out at LSC) and built in collaboration with Alpha Spectra~\cite{AlphaSpectra}. A picture of the set-up is shown in Fig.~\ref{fig:setups}, right, and more details about it can be found in Table~\ref{tab:setups}.

 \end{enumerate}

\begin{table}[h!]
\begin{center}
\caption{Information about the different set-ups in which the studied crystals have been operated: PMT model coupled and if light guides (LG) have been or not used in each detector, and available live time (LT) of data corresponding to each set-up and phase.}
\vspace{0.3cm}
{\begin{tabular}{@{}cccc@{}}
\hline
Set-up& Phase	& PMTs  & LT (days) \\ 		
\hline
1   &  a  &  ETL 9765  &  65.59 \\
    &  b  &  ETL 9765  &  73.49 \\
2   &  I & ETL 9302B (ANAIS-0) & 51.20\\
   &   & ETL 9302B and LG (PIII) & \\
   &  II & Ham. R6233-100MOD (ANAIS-0) & 57.82\\
   &   & ETL 9302B and LG (PIII) & \\
   &  III & Ham. R6956MOD (ANAIS-0) & 39.30\\
   &   & Ham. R11065SEL (PIII) & \\
3   &   & Ham. R6956MOD (ANAIS-25 D0) & 70.40\\
   &   & Ham. R11065SEL (ANAIS-25 D1) & \\
\hline
\end{tabular}\label{tab:setups}}
\end{center}
\end{table}

In all cases, each PMT charge output signal is separately processed; each one is divided into a trigger signal, a signal going to the digitizer, and a number of signals differently attenuated and fed into QDC (charge-to-digital converter) module channels to be integrated in a 1\,${\mu}$s window. At least, a low energy and a high energy ranges were considered in all the analyzed set-ups. The building of the spectra was done by software (off-line) by adding the signals from both PMTs whenever possible.
Data triggering is done in logical OR mode between different detectors operated at the same time and in logical AND between the two PMT signals from each detector, if available (only the BICRON set-up could not use it).

\section{Results and conclusions}
\label{sec:results}

\subsection{$^{40}K$ bulk content}
\label{sec:resultsK}

In this section, we will present the potassium bulk content results for all the studied crystals. The first step in order to derive the $^{40}K$ activity in one crystal, is to select the corresponding energy windows containing the 1460.8\,keV energy depositions in the other(s) crystal(s) sharing the experimental space.
The $^{40}K$ activity of every crystal can be estimated with the area of the 3.2\,keV peak {\it (Area)} identified in the coincident spectra with the high energy window chosen in the other(s) detector(s), the total available live time {\it (t)}, the crystal mass {\it (m)}, the efficiency of the coincidence, determined by simulation using Geant4 package {\it ($\epsilon$)} and the fraction of events effectively selected by the coincidence window chosen {\it (F)}:

\begin {equation}
Activity (Bq/kg) = \frac{Area (counts)}{t(s)\cdot m(kg)\cdot \epsilon \cdot F}
 \label{eq:Activity}
\end {equation}

Only statistical errors coming from the 3.2\,keV peak area determination are taken into account in the derivation of the activity errors shown in the following.

First of all, eleven NaI(Tl) crystals from BICRON, 10.7\,kg mass each were measured: six and seven detectors were placed in a similar configuration in runs {\it a} and {\it b}, respectively, using two of them in both as cross-check. The results for the $^{40}K$ activity corresponding to all of the old BICRON crystals are presented in Table~\ref{tab:PotassiumBICRON}. It can be observed that it is very similar for all of them, ranging from 13 to 21\,mBq/kg, which corresponds to 0.42 to 0.68\,ppm natural potassium in the bulk of the crystals. The procedure followed in order to derive such activity values is equivalent to those explained in detail in the following for ANAIS-0 and PIII crystals, as well as for both ANAIS-25 modules, and is described in detail in section\,\ref{sec:resultsK_a}.

\begin{table}[ht]
\begin{center}
\caption{Results for $^{40}K$ bulk activity of the BICRON\,-\,10.7\,kg NaI(Tl) crystals.}
\vspace{0.3cm}
{\begin{tabular}{@{}cc@{}}
\hline
Detector			& $^{40}K$ Activity (mBq/kg)\\ 	
\hline	
 EP054 & $13.7 \pm 0.3$\\
 EP055 & $15.2 \pm 0.1$\\
 EP056 & $18.8 \pm 0.2$\\
 EP057 & $20.9 \pm 0.4$\\
 EP058 & $16.2 \pm 0.3$\\
 EP059 & $16.6 \pm 0.2$\\
 EL214 & $17.9 \pm 0.4$\\
 EM301 & $21.2 \pm 0.4$\\
 EL604 & $16.5 \pm 0.3$\\
 EL603 & $14.5 \pm 0.2$\\
 EL607 & $15.7 \pm 0.5$\\
\hline
\end{tabular}
\label{tab:PotassiumBICRON}}
\end{center}
\end{table}

\subsubsection{ANAIS-0 and Prototype III.}
\label{sec:resultsK_a}

\begin {figure}[h!]
\centerline{\includegraphics[width=0.95\textwidth]{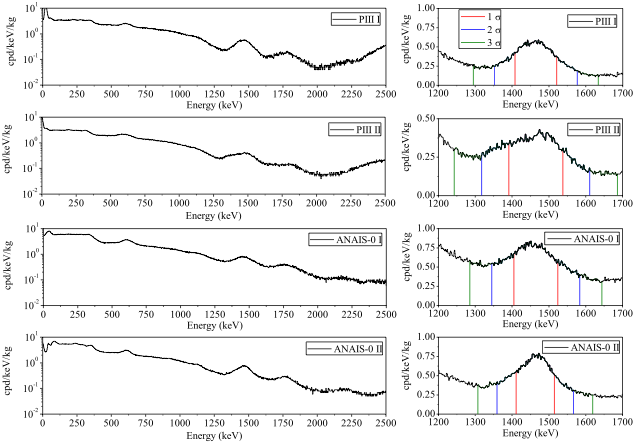}}
\caption{High energy spectra of PIII and ANAIS-0 detectors. Left, the whole spectra and right, a zoom showing the 1460.8\,keV gamma line following $^{40}K$ EC decay for the two considered phases. The $1\,\sigma$ (red), $2\,\sigma$ (blue) and $3\,\sigma$ (green) coincidence windows are also shown.}
\label{fig:40KPIV_HE}
\end {figure}

\begin {figure}[h!]
\centerline{\includegraphics[width=0.95\textwidth]{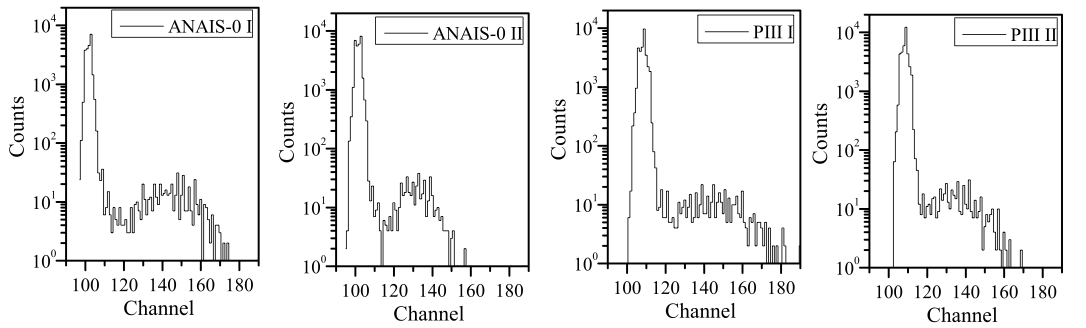}}
\caption{Low energy spectra (in counts/channel) in coincidence with $1\sigma$ windows around the 1460.8\,keV line in the other crystal, shown in Figure~\ref{fig:40KPIV_HE}, for ANAIS-0 and PIII detectors in the two measurement phases considered in this study. The 3.2\,keV peak is clearly visible above the baseline noise peak.}
\label{fig:40KPIV_LE}
\end {figure}

The first step in order to derive the $^{40}K$ activity in both crystals, is to select the corresponding energy windows containing the 1460.8\,keV energy depositions. Different window widths ($1\,\sigma$\footnote{$\sigma$ will refer for each detector to the corresponding value of the standard deviation obtained in the gaussian fit of the 1460.8\,keV line.}, $2\,\sigma$ and $3\,\sigma$) have been considered for the selection of coincident events in both detectors. High energy spectra for the two phases of measurement with ANAIS-0 and PIII detectors are shown in Fig.~\ref{fig:40KPIV_HE}, left, and a zoom around the 1460.8\,keV line, remarking the three windows considered (right). The low energy spectra in coincidence with $1\,\sigma$ window around the 1460.8\,keV line in the other detector are shown in Fig.~\ref{fig:40KPIV_LE}. The gain of PIII along phase II was probably not stable enough and the 1460.8\,keV line is clearly distorted; this could lead to a decrease in the efficiency of the coincidence in an undetermined way, compromising the validity of the derived result.

Only a small fraction of the events selected by the coincidence are attributable to $^{40}K$ decay, but they are, as expected, distributed in a peak around 3.2\,keV, the rest are mostly baseline noise and a few fortuitous coincidences. A simple selection of $^{40}K$ events is done by considering only those events above a threshold. Table~\ref{tab:40KPIVTH} shows the analysis thresholds chosen for each phase and detector. The choice is done just by visual inspection, and it must be remarked that this is not the threshold of the experiment.

\begin{table}[h!]
\begin{center}
\caption{Threshold (in channels) considered for the selection of $^{40}K$ events in the low energy spectra for the two phases for ANAIS-0 and PIII crystals.}
\vspace{0.3cm}
{\begin{tabular}{@{}ccc@{}}
\hline
Detector& Phase & Threshold \\
& & (QDC channel)\\
\hline
ANAIS-0 &I & 115 \\
  &II & 115 \\
PIII &I &  125\\
 &II & 120 \\
\hline
\end{tabular}
\label{tab:40KPIVTH}}
\end{center}
\end{table}

\begin {figure}[h!]
\centering{\includegraphics[width=0.95\textwidth]{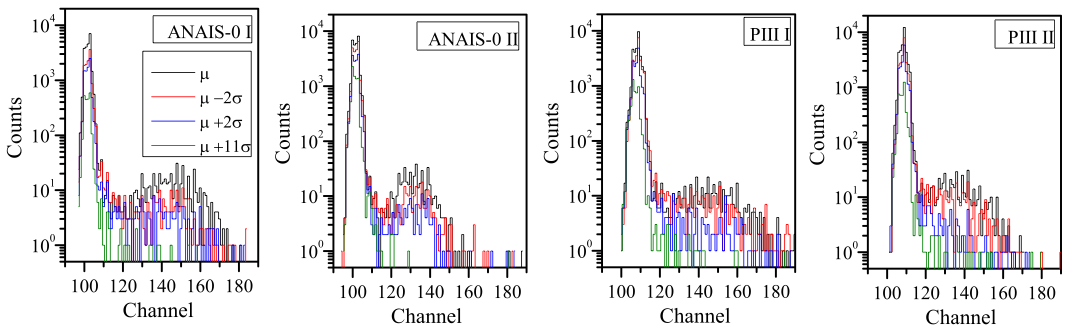}}
\caption{Low energy spectra (in counts/channel) in coincidence with high energy windows in the other detector (all of them having the same width, $\pm\sigma$) are shown for the different phases and detectors: centered at 1460.8\,keV peak ($\mu$) in black, centered $2\sigma$ above ($\mu+2\sigma$) in blue, centered $2\sigma$ below ($\mu-2\sigma$) in red, and centered further above ($\mu+11\sigma$) in green.}
\label{fig:40KPIV_windows}
\end {figure}

In order to check the $^{40}K$ origin of the low energy events selected, as previously explained, the effect of changing the high energy window above and below the 1460.8\,keV position has been studied. For this purpose, the coincidence is done with $1\,\sigma$ width windows centered in channels $2\,\sigma$ above and below 1460.8\,keV. Also another window more energetic is selected, centered $11\,\sigma$ above the 1460.8\,keV position (see results in Fig.~\ref{fig:40KPIV_windows}). Assuming gaussian shape for the 1460.8\,keV line, the corresponding percentage of real $^{40}K$ events selected in each window should be 68\% for $\mu\pm\sigma$, 16\% for $\mu+2\sigma\pm\sigma$ and 0 for the $\mu+11\sigma\pm\sigma$. The highest window should only present fortuitous coincidences and could allow us to estimate their contribution in the other windows. In Table~\ref{tab:windows} the results are presented considering only events above the thresholds shown in Table~\ref{tab:40KPIVTH}. As expected, events at low energy coincident with the $\mu+11 \sigma$ window correspond to fortuitous coincidences, and the peak is not seen. The $\mu +2\sigma$ window presents results compatible with the expected 16\% of the total number of coincident events. Events found in coincidence with the $\mu -2 \sigma$ window are much more than expected for a pure gaussian peak contribution, but this is probably due to the presence of multi-Compton events with partial energy deposition from the $^{40}K$ gamma line. Numbers shown in Table~\ref{tab:windows} do not allow directly estimate of the fortuitous coincidence rate contribution, because this is strongly related to the total rate in the high energy coincidence window (much lower for instance in the $\mu+11 \sigma$ window); then, conclusions derived from this table are only qualitatively valid.

\begin{table}[h!]
\begin{center}
\caption{Number of events selected by the coincidence above the thresholds, shown in Table~\ref{tab:40KPIVTH}, in the different windows studied (all of them having the same width $\pm\sigma$). Effective exposure corresponding to every phase is given in Table~\ref{tab:setups}.}
\vspace{0.3cm}
{\begin{tabular}{@{}cccccc@{}}
\hline
Detector & {Phase}&	$\mu$ & $\mu +2 \sigma$ & $\mu -2 \sigma$& $\mu + 11 \sigma$\\
&&\multicolumn{4}{c}{Events}\\
\hline
ANAIS-0 &I & 726   & 95& 312 & 8\\
 & II & 713  & 142& 253 & 26\\
PIII &I & 706  & 146 & 358 & 22\\
 &II & 766  & 137& 205 & 12\\
\hline
\end{tabular}\label{tab:windows}}
\end{center}
\end{table}

Then, the set-up has been simulated with Geant4, version geant4.9.1.p02 \cite{Geant4}, in order to evaluate the probability that, after a $^{40}K$ disintegration in one crystal, the 1460.8\,keV photon escapes and releases the full energy in the other detector\footnote{This simulation has been specifically done for every experimental set-up.}. 500000\,photons of 1460.8\,keV have been simulated assuming homogeneous distribution of the contaminant in the bulk in ANAIS-0 and PIII crystals. The absolute branching ratio for the $^{40}K$ K-shell EC followed by the emission of the 1460.8\,keV photon is 0.0803, as given by Geant4~\cite{Geant4}. The number of events with the full gamma energy absorbed in PIII or ANAIS-0 crystals when emitted in ANAIS-0 and PIII are 8258 and 6998, respectively. Thus, the efficiencies for the observation of the respective coincidences per $^{40}K$ decay are $1.33\cdot10^{-3}$ and $1.13\cdot10^{-3}$.

\begin {figure}[h!]
\centering{\includegraphics[width=0.8\textwidth]{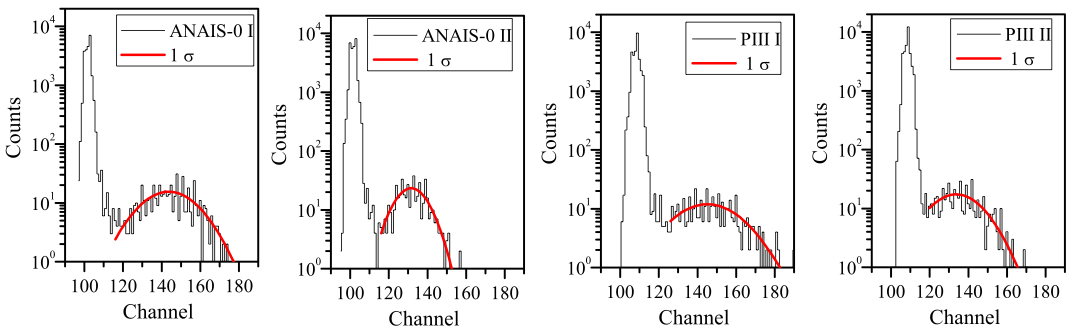}
\includegraphics[width=0.8\textwidth]{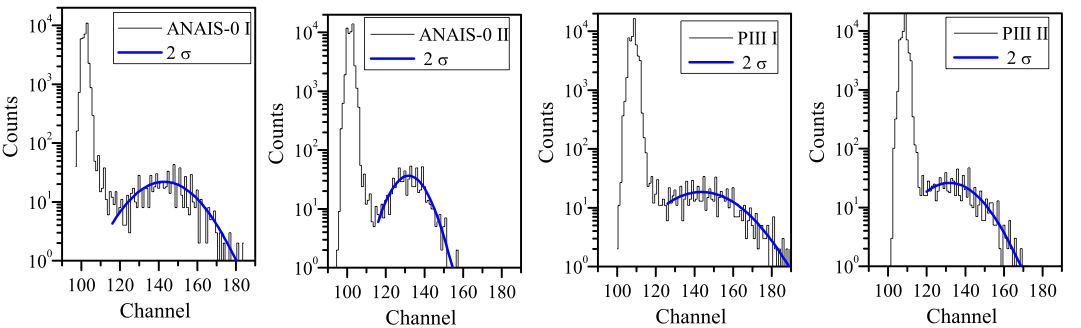}
\includegraphics[width=0.8\textwidth]{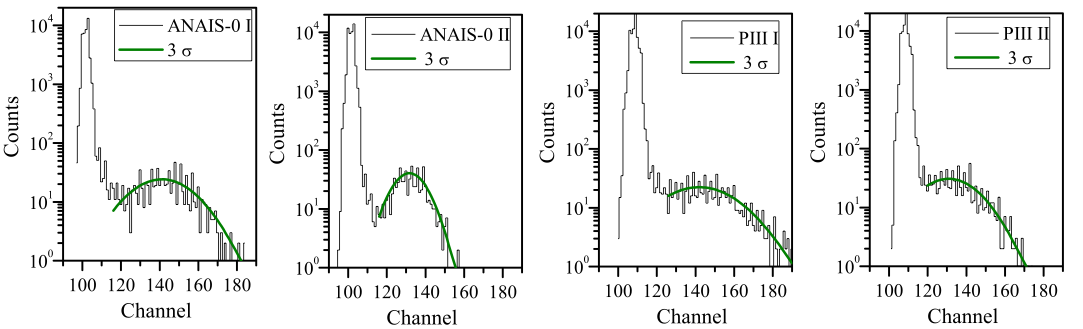}}
\caption{Low energy coincident events (in counts/channel) for the $1\sigma$, $2\sigma$ and $3\sigma$ coincidence windows, and gaussian fits of the events above the threshold.}
\label{fig:40KPIV_fit}
\end {figure}

The area of the 3.2\,keV peak (Area) is obtained by fitting to a gaussian the events above the threshold. The fits are shown in Fig.~\ref{fig:40KPIV_fit}. The $^{40}K$ activity is calculated following eq.\,\ref{eq:Activity} for each phase individually and using all the available data for the three different coincidence window widths ($1\sigma$, $2\sigma$ and $3\sigma$) around 1460.8\,keV energy. Results for each phase and detector are shown in Table~\ref{tab:fits}.

\begin{table}[h!]
\begin{center}
\caption{$^{40}K$ activity calculated for ANAIS-0 and PIII using different width coincidence windows. Combined values derived from the first two phases for each detector are also shown. }
\vspace{0.3cm}
{\begin{tabular}{@{}ccccc@{}}
\hline
Detector & Phase &\multicolumn{3}{c}{$^{40}K$ Activity (mBq/kg)}\\
				&				&$1\,\normalfont\sigma$ & $2\,\normalfont\sigma$ & $3\,\normalfont\sigma$ \\
\hline
ANAIS-0  &I & $14.3 \pm 0.8$ & $15.1  \pm 0.9 $& $17.2  \pm 1.1$\\
	        &II & $11.1\pm 0.5$ & $12.4 \pm 0.5 $& $13.4  \pm 0.6 $\\
	        &I and II &$ 12.7 \pm 0.5$ & $13.6  \pm 0.5$ &$15.2 \pm 0.6 $\\
	        &III & $14.5 \pm 0.6$ & $14.3 \pm 0.5 $& $15.1 \pm 0.5$\\
\hline
PIII  &I& $13.5 \pm 0.9$ & $16.8\pm 1.13 $& $20.1 \pm 1.4 $\\
       &II&$13.9\pm 0.9$ & $16.1  \pm 1.1 $& $19.0  \pm 1.4 $\\
       &I and II&$ 13.7 \pm 0.6$ & $16.4  \pm 0.7$&$19.5\pm 1.0 $\\
\hline
\end{tabular} \label{tab:fits}}
\end{center}
\end{table}

Results derived for the different windows and phases analyzed are mostly compatible. As expected, larger windows have a larger contribution from fortuitous coincidences and Compton events. Hence, results of the $1\,\sigma$ window have been taken in the following as the most reliable. Phase II presented a non gaussian shape for the 1460.8\,keV gamma line of PIII which might have been caused by gain instabilities. However, phase I showed a higher discrepancy between the results determined with the different sigma windows and in Fig.~\ref{fig:40KPIV_fit} it can be seen that the 3.2\,keV peak is wider in phase I and more contribution from fortuitous coincidences is expected. In PIII both phases gave similar results. Then, the average of phase I and II results for every detector was taken as final result of our analysis and used, for instance in Ref.\,\citen{ANAISbkg}.

\begin {figure}[h!]
\centering{\includegraphics[width=0.8\textwidth]{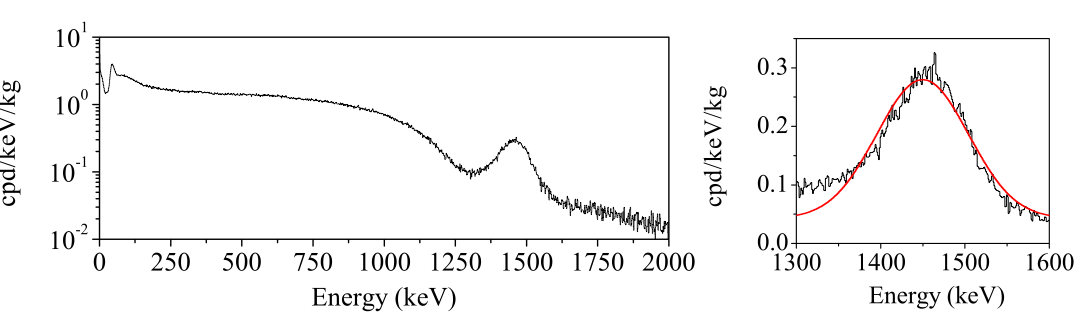}}
\caption{High energy spectrum for the ANAIS-0 module used to derive the $^{40}K$ bulk content from the intensity of the background line at 1461-1464\,keV, assuming negligible contributions from external $^{40}K$ sources.}
\label{fig:A02_40K}
\end {figure}

We checked that these results were compatible with the intensity of the 1461-1464\,keV gamma line seen at ANAIS-0 background. Bulk crystal potassium contamination contributes to this line with different energy depositions: from K-shell EC decay of $^{40}K$, producing 1464.0\,keV (1460.8\,keV\,+\,3.2\,keV) total energy release, but also from L and M-shell EC decays, with energy depositions that can not be distinguished from the 1460.8\,keV line. Moreover, any other external $^{40}K$ contamination would also contribute to the 1460.8\,keV line. We chose data from ANAIS-0 operating without PIII to minimize contributions from external components contaminated in $^{40}K$ to the photopeak and fitted it to a gaussian, comparing its area with the prediction of our Geant4 simulation (see Fig.\,\ref{fig:A02_40K}). For 500000 isotropic 1460.8\,keV photons simulated, the ANAIS-0 crystal detects 135294 photons in the photopeak (27.1\%). Taking into account that only in 10.55\% of the $^{40}K$ decays a high energy gamma is emitted, if the result for this 1461-1464\,keV peak is $32.38\pm0.62$\,cpd/kg, the activity of $^{40}K$ derived is $13.11\pm0.25$\,mBq/kg assuming that only $^{40}K$ in the crystal bulk is contributing. This result is compatible with the activity derived from the coincidence measurement and implies that the ANAIS-0 background is dominated by $^{40}K$ in the bulk, as confirmed the background model proposed and simulated in Ref.\,\citen{ANAISbkg}.

\begin {figure}[h!]
\centering{\includegraphics[width=0.8\textwidth]{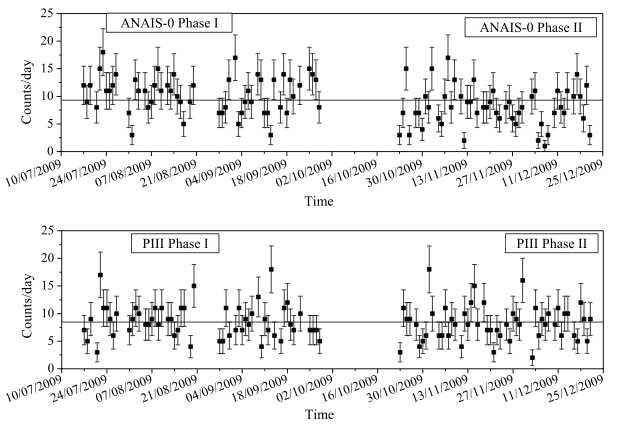}}
\caption{Rate of the 3.2\,keV events selected by the coincidence above the threshold. Average values of $9.3\pm3.6$\,counts/day for ANAIS-0 and of $8.5\pm3.1$\,counts/day for PIII crystals are shown as horizontal lines in both plots.}
\label{fig:40KPIV_rate}
\end {figure}

At last, the temporal distribution of $^{40}K$ events at low energy selected by the coincidence above the threshold is shown in Fig.~\ref{fig:40KPIV_rate}. Average values of $9.3\pm3.6$\,counts/day for ANAIS-0 and of $8.5\pm3.1$\,counts/day for PIII crystals are reported without significant fluctuations.

\begin {figure}[h!]
\centering{\includegraphics[width=0.8\textwidth]{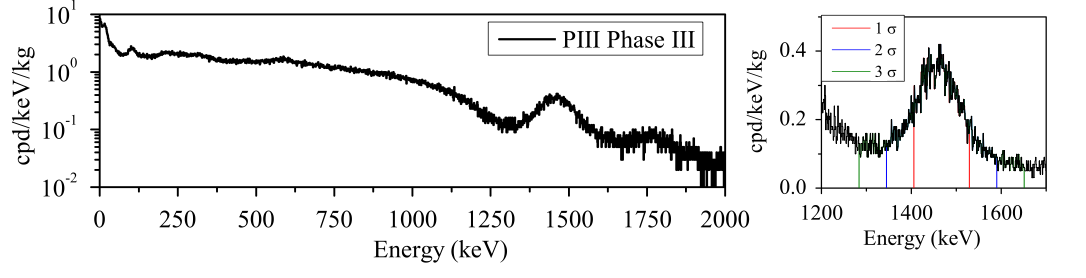}
\includegraphics[width=0.8\textwidth]{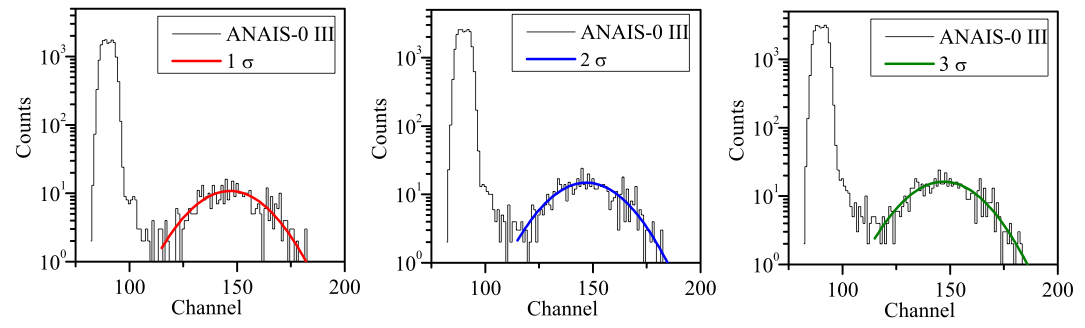}}
\caption{Top: High energy spectrum of PIII corresponding to the phase III of the $^{40}K$-coincidence set-up (left) and zoom showing the 1460.8\,keV line and the $1\,\sigma$ (red), $2\,\sigma$ (blue) and $3\,\sigma$ (green) coincidence windows (right). Bottom: Low energy coincident events (in counts/channel) for the $1\sigma$, $2\sigma$, and $3\sigma$ coincidence windows, and gaussian fits of the events above the threshold.}
\label{fig:40KPhIII}
\end {figure}

After this analysis was completed, a new estimate of the $^{40}K$ content of ANAIS-0 crystal was derived using data from the phase III $^{40}K$ coincidence set-up. Corresponding results are shown in Fig.~\ref{fig:40KPhIII} and Table~\ref{tab:fits}. Very nice and stable operation can be reported and results are compatible with those of phase I, slightly higher than those of phase II, and pointing at some coincident events loss in phase II attributable to instability in PIII high energy data. Our conclusion is that the $^{40}K$ activity we assumed for ANAIS-0 (derived from the coincident 3.2\,keV peak intensity, by averaging the estimates from phase I and II) is underestimated in about a 10\%. The limit for $^{40}K$ derived from the 1460.8\,keV gamma line in the background is more hardly compatible with such a higher $^{40}K$ bulk content, but systematics on Geant4 simulations could be responsible of such an underestimate: for instance, possible energy loss mechanisms could affect in about such a percent the conclusions derived from our analysis.

\subsubsection{ANAIS-25}
\label{sec:resultsK_b}

\begin {figure}[h]
\centering{\includegraphics[width=0.8\textwidth]{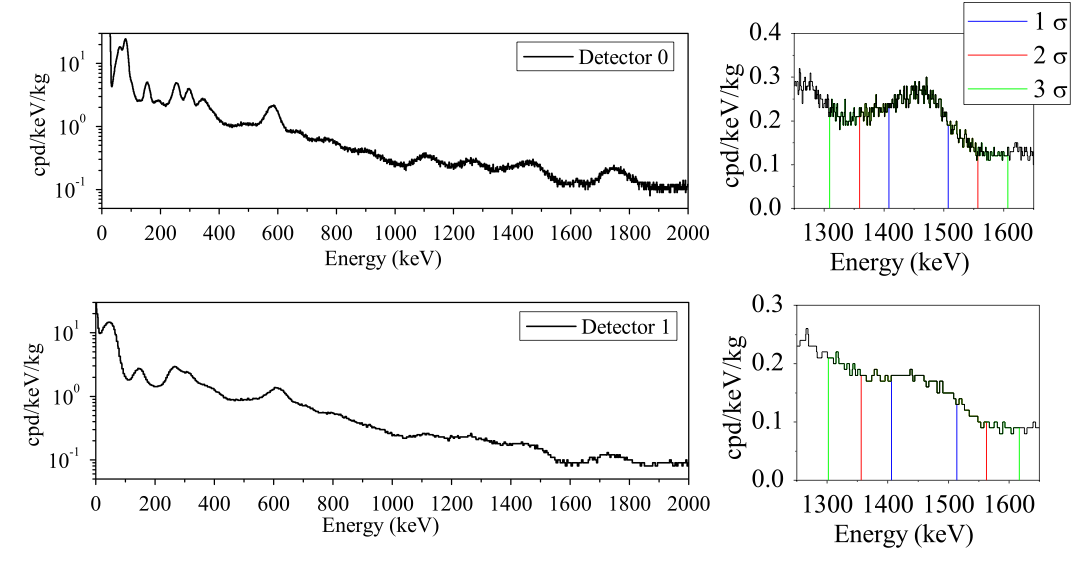}}
\caption{High energy spectra of the ANAIS-25 modules corresponding to 70.4\,days (left), and zoom showing the 1460.8\,keV gamma line in the right. The $1\,\sigma$ (red), $2\,\sigma$ (blue) and $3\,\sigma$ (green) coincidence windows are also shown.}
\label{fig:A25_40K_HE}
\end {figure}

\begin {figure}[h]
\centering{\includegraphics[width=0.8\textwidth]{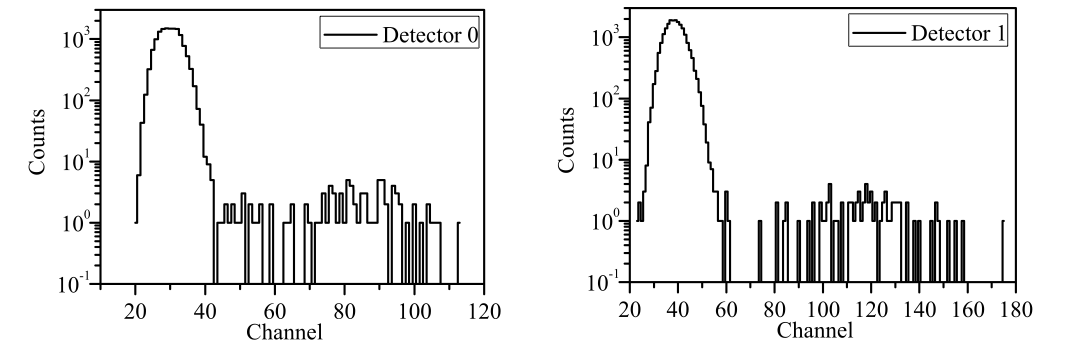}}
\caption{Low energy spectra (in counts/channel) in coincidence with $1\,\sigma$ windows around the 1460.8\,keV line (shown in Figure~\ref{fig:A25_40K_HE}) in the other crystal for ANAIS-25 D0 (left) and D1 (right).}
\label{fig:A25_40K_LE}
\end {figure}

\begin {figure}[h]
\centering {\includegraphics[width=0.8\textwidth]{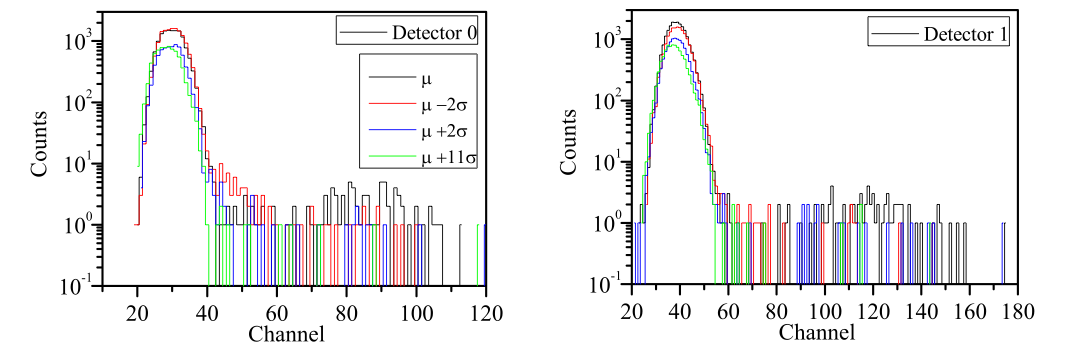}}
\caption{Low energy spectra (in counts/channel) in coincidence with high energy windows in the other detector (all of them having the same width, $\pm\sigma$) are shown for ANAIS-25 D0 (left) and D1 (right): centered at 1460.8\,keV peak ($\mu$) in black, centered $2\sigma$ above ($\mu+2\sigma$) in blue, centered $2\sigma$ below ($\mu-2\sigma$) in red, and centered further above ($\mu+11\sigma$) in green.}
\label{fig:A25_40K_LEw}
\end {figure}

The potassium content of the ANAIS-25 modules has been carefully analyzed directly applying the same technique. The $^{40}K$ gamma events at 1460.8\,keV in one detector are selected considering different windows widths ($1\,\sigma$, $2\,\sigma$ and $3\,\sigma$) as done with ANAIS-0 and PIII data. High energy spectra of both detectors are shown in Fig.~\ref{fig:A25_40K_HE}. The bad resolution observed is a consequence of the instabilities in PMTs gain, specially in detector~1 where only data from PMT~1 have been considered for the first weeks of data, because a fast estimate of the potassium content was important and no data were discarded. Excellent performance of the set-up has been demonstrated later on. The low energy spectra in coincidence with the $1\,\sigma$ window around the 1460.8\,keV line in the other detector, are shown in Fig.~\ref{fig:A25_40K_LE}. The threshold chosen to select the events attributable to $^{40}K$ decay are channel~60 for D0 and channel~70 for the D1. The effect of changing the high energy window above and below the 1460.8\,keV position has also been studied, see results in Figure~\ref{fig:A25_40K_LEw}. Numbers of counts over these thresholds for every coincidence window are shown in Table~\ref{tab:A25wcts}.

\begin{table}[h]
\begin{center}
\caption{Events above the analysis threshold (channel 60 for D0 and channel 70 for D1), among those selected by the coincidence with an event in a window centered at, below, or above the 1460.8\,keV gamma position (1\,$\sigma$ width).}
\vspace{0.3cm}
{\begin{tabular}{@{}ccccc@{}}
\hline
Detector& $\mu$&$\mu\,-2\,\sigma$  & $\mu\,+2\,\sigma$&$\mu\,+11\,\sigma$ \\
\hline
0 & 82 & 25 & 16& 7\\
1& 78 & 22 & 21&10\\
\hline
\end{tabular}
\label{tab:A25wcts}}
\end{center}
\end{table}

\begin {figure}[h!]
\centering{\includegraphics[width=0.8\textwidth]{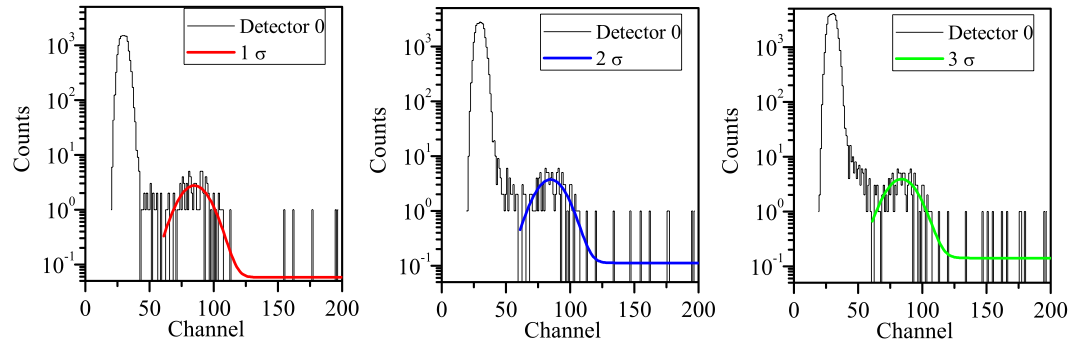}
\includegraphics[width=0.8\textwidth]{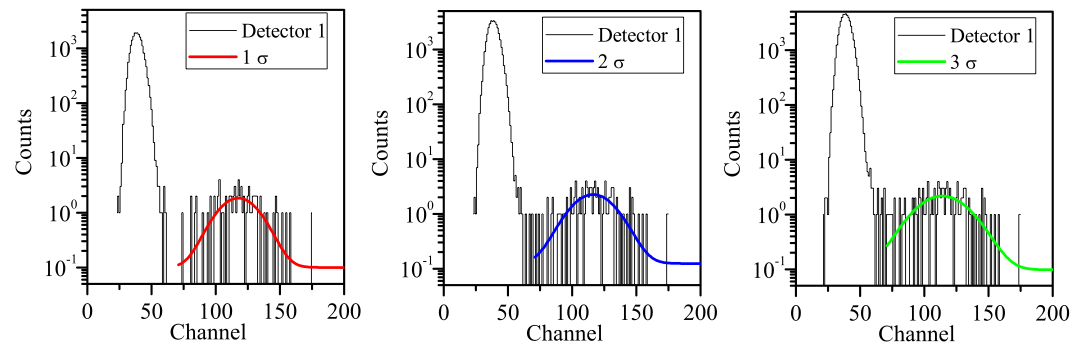}}
\caption{Low energy spectra (in counts/channel) in coincidence with $1\,\sigma$, $2\,\sigma$, and $3\,\sigma$ windows around 1460.8\,keV line in the other crystal for ANAIS-25 D0 and D1.}
\label{fig:A25_40K_LEfit}
\end {figure}

\begin {figure}[h!]
\centering{\includegraphics[width=0.5\textwidth]{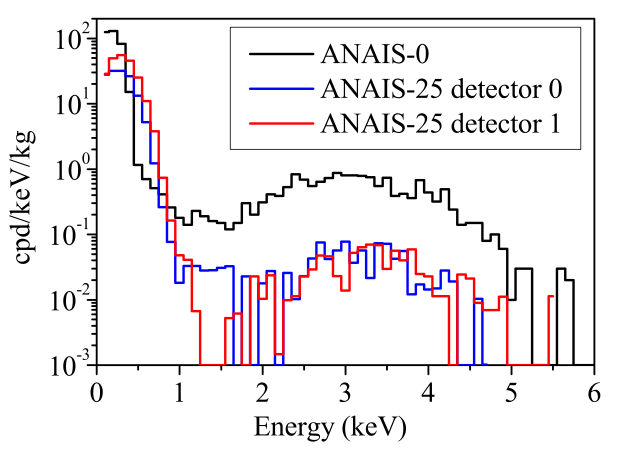}}
\caption{Low energy spectra in coincidence with $1\,\sigma$ windows around 1460.8\,keV line in the other crystal for ANAIS-0 (black), ANAIS-25 detector 0 (blue), and ANAIS-25 detector 1 (red).}
\label{fig:A25_40K_LEcn}
\end {figure}

The probability that, after a $^{40}K$ disintegration in one crystal, the 1460.8\,keV photon escapes and releases the full energy in the other detector has been estimated with Geant4, in this case using version geant4.9.4.p01. The corresponding efficiency for the coincidences between both ANAIS-25 modules is $1.08\cdot10^{-3}$, just a bit lower than that obtained for ANAIS-0 and PIII because of the higher mass, that decreases the probability for the escape of the gamma without losing any energy. Then, the activities of $^{40}K$ for each ANAIS-25 crystal have been estimated for the different width coincidence windows, see equation \ref{eq:Activity}. The gaussian fits performed to redeem the area are shown in Fig.~\ref{fig:A25_40K_LEfit}. Results on the $^{40}K$ content for each detector are shown in Table~\ref{tab:A25_40KA}. Good agreement between results derived for both detectors is observed, as expected. Averaging the  $1\,\sigma$ window results for the two crystals, we can conclude that ANAIS-25 crystals have a $^{40}K$ content $1.25\pm0.11$\,mBq/kg ($41.7\pm3.7$\,ppb of potassium) much lower than that estimated for ANAIS-0 crystal, see Fig.~\ref{fig:A25_40K_LEcn}.

\begin{table}[h!]
\begin{center}
\caption{$^{40}K$ Activity calculated for the two ANAIS-25 crystals using different widths coincidence windows.}
\vspace{0.3cm}
{\begin{tabular}{@{}cccc@{}}
\hline
Detector	&\multicolumn{3}{c}{$^{40}K$ Activity (mBq/kg)}\\
					& $1\,\sigma$		&$2\,\sigma$		&$3\,\sigma$  \\
\hline
0				 	& $1.34\pm0.13$ & $1.16\pm0.11$	&$1.31\pm0.11$\\
1					& $1.15\pm0.18$ & $1.08\pm0.16$	&$1.21\pm0.20$\\
\hline
\end{tabular}
\label{tab:A25_40KA}}
\end{center}
\end{table}

NaI(Tl) crystals from different manufacturers have been characterized in terms of their potassium bulk content by a measurement in coincidence, and improvement of one order of magnitude in the potassium content can be reported for the ANAIS-25 detectors, built in collaboration with Alpha Spectra. However, the 20\,ppb goal has not yet been achieved and before ordering the additional 18 modules required to complete the ANAIS total detection mass, careful analysis of the situation in collaboration with Alpha Spectra is undergoing, trying to further purify the starting NaI powder.

\subsection{Trigger efficiency at 3.2\,keV}

\begin {figure}[h!]
\centering{\includegraphics[width=0.8\textwidth]{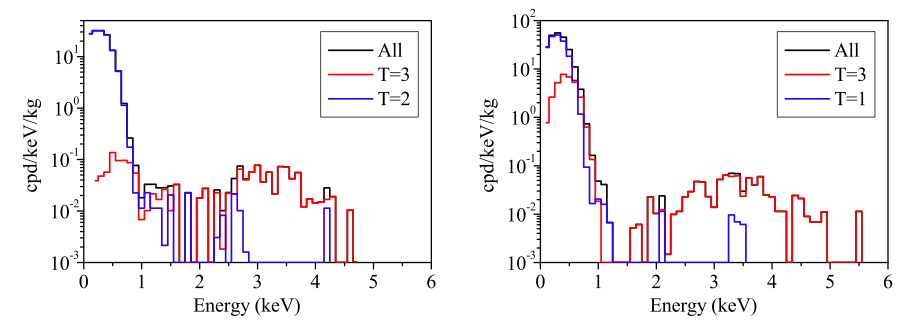}}
\caption{$^{40}K$ events at low energy, identified by the coincidence with a high energy gamma for the ANAIS-25 D0 (left), and D1 (right). Events with T=2 (T=1) have not triggered in D0 (D1), whereas events with T=3, have triggered in both detectors.}
\label{fig:trigger_eff}
\end {figure}

However, having a tagged population of bulk scintillation events at 3.2\,keV is very useful for many other purposes related to the DM search. Just to show an example, in ANAIS-25 setup we used this $^{40}K$ events to estimate the trigger efficiency of the experiment. It is shown in Fig.\,\ref{fig:trigger_eff} how many of the low energy events identified by the coincidence with the high energy window around 1460.8\,keV, and hence, corresponding to the decay of $^{40}K$, have effectively triggered our acquisition. A good trigger efficiency can be reported: 99\% of the events above 1.5\,keV are triggering in D1, and 97\% in D0. It is worth noting that fortuitous coincidences in D0 arrive up to higher energies (which is related to the higher dark current of the PMTs used), and that in D1, baseline is eventually triggering.

\section*{Acknowledgments}

This work has been supported by the Spanish Ministerio de Econom\'{\i}a y Competitividad and the European Regional Development Fund (MINECO-FEDER) (FPA2011-23749), the Consolider-Ingenio 2010 Programme under grants MULTIDARK CSD2009- 00064 and CPAN CSD2007-00042, and the Gobierno de Arag\'{o}n (Group in Nuclear and Astroparticle Physics, ARAID Foundation and C. Cuesta predoctoral grant). C. Ginestra and P. Villar have been supported by the MINECO Subprograma de Formaci\'{o}n de Personal Investigador. We also acknowledge LSC and GIFNA staff for their support.


\begin{thebibliography}{00}

\bibitem{DAMA} R. Bernabei, {\it et al.}, {\it Riv. Nuovo Cim. A} {\bf 112}, 545 (1999).\\
R.~Bernabei, {\it et al.}, {\it Phys. Lett. B} {\bf 450} 448 (1999).
\bibitem{LIBRA} R. Bernabei, {\it et al.}, {\it Nucl. Inst. Meth. A} {\bf 592}, 297 (2008).\\
R.~Bernabei, {\it et al.}, {\it Eur. Phys. J. C} {\bf 56} 333 (2008). \\
R. Bernabei, {\it et al.}, {\it Eur. Phys. J. C} {\bf 67}, 39 (2010).\\
R.~Bernabei, {\it et al.}, {\it Nucl. Instr. Meth. A} {\bf 692} 120 (2012).\\
R. Bernabei, {\it et al.}, {\it Eur. Phys. J. C} {\bf 73} 2648 (2013).
\bibitem{DM32} M.L. Sarsa, {\it et al.}, {\it Nucl. Phys. B (Proc. Suppl.)} {\bf 35}, 154 (1994).\\
M.L. Sarsa, {\it et al.}, {\it Phys. Lett. B} {\bf 386}, 458 (1996).\\
M.L. Sarsa, {\it et al.}, {\it Phys. Rev. D} {\bf 56}, 185 (1997).
\bibitem{JAPAN1}
K.~Fushimi {\it et al.}, {\it Phys. Rev. C} {\bf 47} 425 (1993). \\
K.~Fushimi {\it et al.}, {\it Astropart. Phys.} {\bf 12} 185 (1999).
\bibitem{PSD_Gerbier}
G.~Gerbier {\it et al.}, {\it Astropart. Phys.} {\bf 11} 287 (1999).
\bibitem{UKDMS}
G.J.~Alner {\it et al.}, {\it Phys. Lett. B} {\bf 616} 17 (2005).
\bibitem{JAPAN2}
K.~Fushimi {\it et al.}, {\it J. Phys.: Conf. Ser.} {\bf 203} 012046 (2010).
\bibitem{DM-ice}
J.~Cherwinka {\em et al.}, {\it Astropart. Phys.} {\bf 35} 749 (2012).
\bibitem{modanedobleb2} A. Morales, {\it et al.}, {\it Il Nuovo Cimento A} {\bf 100}, 525 (1988).\\
\bibitem{moralesdobleb2} A. Morales, {\it et al.}, {\it J. Phys. G (Nucl. Phys.)} {\bf 17}, 211 (1991).\\
A. Morales, {\it et al.}, {\it Il Nuovo Cimento A} {\bf 104}, 1581 (1991).
\bibitem{tesisSusana} S. Cebri\'an, {\it Estudio del fondo radiactivo en experimentos subterr\'aneos de b\'usqueda de sucesos poco probables: CUORE (Cryogenic Underground Observatory for Rare Events) y ANAIS (Annual modulation with NaI(Tl))}, Ph.D. Dissertation, Universidad de Zaragoza, 2002.
\bibitem{tesisMaria} M. Mart\'{\i}nez, {\it Dise\~no de un prototipo para un experimento de detecci\'on directa de materia oscura mediante modulaci\'on anual con centelleadores de ioduro de sodio}, Ph.D. Dissertation, Universidad de Zaragoza, 2006.
\bibitem{SaintGobain} Saint Gobain (\url{http://www.saint-gobain.com/en})
\bibitem{TesisClara} C. Cuesta, {\it ANAIS-0: Feasibility study for a 250\,kg NaI(Tl) dark matter search experiment at the Canfranc Underground Laboratory}. Ph.D. Dissertation, Universidad de Zaragoza, 2013.
\bibitem{ANAISbkg} S. Cebri\'an, {\it et al.}, {\it Astrop. Phys.} {\bf 37}, 60 (2012).
\bibitem{AlphaSpectra} Alpha Spectra Inc., Grand Junction, Colorado, US. \url{http://www.alphaspectra.com/}
\bibitem{Pradler} J. Pradler, B. Singh, I. Yavin, {\it Phys. Lett. B} {\bf 720}, 399 (2013).
\bibitem{nucleide} \url{http://www.nucleide.org}
\bibitem{ANAISom} C. Cuesta, {\it et al.}, {\it Opt. Mat.} {\bf 36}, 316 (2013).
\bibitem{Geant4} S. Agostinelli, {\it et al.}, {\it Nucl. Inst. Meth. A} {\bf 506}, 250 (2003).

\end{thebibliography}
\end{document}